\documentclass{iau}
\usepackage{graphicx} 

\title[IAUS291.~~Galactic Millisecond Pulsars] 
{The Galactic Millisecond Pulsar Population} 

\author[Duncan R. Lorimer]  
{Duncan R. Lorimer$^{1,2}$}
\affiliation{$^1$Department of Physics, West Virginia University, USA\\
$^2$National Radio Astronomy Observatory, Green Bank, USA
\\email: {\tt Duncan.Lorimer@mail.wvu.edu}}
\pubyear{2012}
\volume{291}  
\jname{\mbox{Neutron Stars and Pulsars: Challenges and Opportunities after 80 years}}
\editors{J.~van Leeuwen, ed.} 
\begin{document}

\maketitle

\begin{abstract}

Among the current sample of over 2000 radio pulsars known primarily in
the disk of our Galaxy, millisecond pulsars now number almost
200\footnote{For my list of Galactic millisecond
  pulsars, see http://astro.phys.wvu.edu/GalacticMSPs}. Due
to the phenomenal success of blind surveys of the Galactic field, and
targeted searches of {\it Fermi} gamma-ray sources, for the first time
in over a decade, Galactic millisecond pulsars now outnumber their
counterparts in globular clusters! In this paper, I briefly review
earlier results from studies of the Galactic millisecond pulsar
population and present new constraints based on a sample of 60
millisecond pulsars discovered by 20 cm Parkes multibeam surveys. I
present a simple model of the population containing $\sim 30,000$
potentially observable millisecond pulsars with a luminosity function,
radial distribution and scale height that matches the observed sample
of objects. This study represents only a first step towards a more
complete understanding of the parent population of millisecond pulsars
in the Galaxy and I conclude with some suggestions for further study
in this area.

\end{abstract}


\firstsection 
\section{Introduction}

Millisecond pulsars have been the subject of intense discovery over
the past few years. Thanks to the current generation of large-scale
pulsar surveys, we find ourselves in an era where we have large
samples of both millisecond and normal pulsars and, in particular for
millisecond pulsars, there are many opportunities to learn about the
population of objects as a whole based upon the ones we see. The
contributions by Keith, Ng, Lazarus and Lynch elsewhere in these
proceedings provide the latest status of these searches, and further
details can be found in Lorimer (2011). 
These discoveries are due to the modern surveys having low-noise
receiver systems with a large fractional bandwidth and employing
state-of-the-art digital data acquisition systems which are now close
to optimal (e.g.~DuPlain et al.~2008), as well as substantial
computing resources using sophisticated search processing algorithms
which can largely remove radio-frequency interference and combat the
effects of binary motion during the observation.  In addition,
multiple analyses of the data often result in additional discoveries
(e.g. Keith et al.~2009; Eatough et al.~2010; Mickaliger et al.~2012).

Along with the current generation of ongoing pulsar surveys at Green
Bank, Parkes, Arecibo and Effelsberg, a key set of surveys that
provide the backbone for much of the current population analyses are
the Parkes multibeam surveys. These groundbreaking experiments were
carried out using analog filterbanks and had substantially better
sensitivity to millisecond pulsars compared to previous efforts. The
main surveys of interest here are the Galactic plane survey
(Manchester et al. 2001), the Swinburne intermediate (Edwards et
al.~2001) and high-latitude (Jacoby et al.~2009) pulsar surveys, the
high latitude survey (Burgay et al.~2006) and Perseus arm survey
(Burgay et al.~2012) as well as a deep multibeam survey of the
northern Galactic plane (Lorimer, Camilo \& McLaughlin 2013).  In
spite of their good sensitivity, these surveys were ultimately limited
by a number of selection effects which bias their pulsar samples and
need to be accounted for by population analyses. These selection
effects include the inverse-square law, pulse dispersion and
scattering, pulsar intermittency, interstellar scintillation and also
binary motion.  For further details of these effects, the interested
reader is referred to earlier reviews on this subject (Lorimer 2009, 2011).

\section{Overiew of modeling approaches}

Although early efforts to correct for observational selection were
done via analytical treatments (e.g.~Gunn \& Ostriker 1970), nowadays
this is best done in a Monte Carlo fashion to create realizations of
the true of the underlying pulsar population which are then searched
using models of survey detection thresholds which account for
propagation in the interstellar medium. The contribution by Cordes in
this volume describes current work to update the so-called ``NE2001''
electron density model (Cordes \& Lazio 2002) which is currently used
to model the propagation effects in these population syntheses. The reader
is also referred to recent work on this subject by Schnitzeler (2012).

From statistical analyses of samples of artificial pulsars that
satisfy the criteria for detection, it is possible to generate and
optimize models for the pulsar population which inform us about the
underlying distribution functions and make predictions for future
survey yields. Although a variety of different approaches have been
employed, the Monte Carlo simulations follow two basic strategies. In
the ``snapshot'' approach, no assumptions are made concerning the
prior evolution of pulsars. Instead, the populations are simply
generated according to various distribution functions (typically in
Galactocentric radius, $R$, height with respect to the plane, $z$,
spin period, $P$ and luminosity, $L$) which are optimized in order to
find the best match to the sample. Alternatively, one may carry out
``evolution'' approaches where the model pulsars are evolved forward
in time from a set of initial distributions. Software to carry out
both approaches has been developed by a number of groups and some of
this is freely available\footnote{For example, the {\scshape psrpop}
  software package at http://psrpop.phys.wvu.edu has modules to carry
  out both the snapshot and evolution approaches}.

The snapshot approach was applied to the normal pulsar population by
Lorimer et al.~(2006) who were able to derive best-fitting probability
density functions in $R$, $L$, $z$ and $P$ for the present-day
population of objects. One result of this work was that the radial
distribution of pulsars could not be decoupled from the radial
distribution of free electrons in the pulsar distribution.  For the
evolution approach on the normal population, the current
state-of-the-art is the work of Faucher-Gigu\`ere \& Kaspi (2006) who
generated excellent fits to the pulsar $P$-$\dot{P}$ diagram using a
model in which the luminosity has a power-law dependence on $P$ and
$\dot{P}$. In their optimal model, $L$ scales as $P^{-1.5}
\dot{P}^{0.5}$, i.e.~the square root of the spin-down luminosity. One
interesting result from this paper is that the luminosity function of
the present-day pulsar population appears to be log-normal in
form. The smooth tails of this distribution (which are integrable over
all luminosities to give a finite result) offer a distinct advantage
over previous studies which parameterized the luminosity as a power
law which is divergent and requires a somewhat unphysical minimum
luminosity. Similar results were found by Ridley \& Lorimer
(2010). The log-normal form of the luminosity distribution has
subsequently been adopted as a starting point by a number of other
studies (e.g., Boyles et al. 2011; Bagchi et al. 2011 and
Chennamangalam et al.~2012; see also these proceedings).

\section{Previous studies of the millisecond pulsar population}

One of the first efforts to quantify the millisecond pulsar population
was the work of Kulkarni \& Narayan (1988) who used
a $V/V_{\rm max}$ approach to estimate the number of similar objects
to those observed by surveys at that time. With a sample of only three
millisecond pulsars, their study was subject to large uncertainties,
but it began a significant discussion on the so-called ``birthrate
problem'' for millisecond pulsars. Based on their results Kulkarni \&
Narayan (1989) claimed that the birthrate of millisecond pulsars was
substantially greater than that of their proposed progenitors, the
low-mass X-ray binaries. This problem has largely disappeared as
better constraints have become available from larger samples (Lorimer
2009). Rathnasree (1993) attempted to synthesize the population of
millisecond pulsars from low-mass X-ray binaries by carrying out Monte
Carlo simulations to model their evolution since birth. The current
state of the art of this approach is discussed in the contribution by
Tauris in this proceedings.

A prescient paper by Johnston \& Bailes (1991) 
demonstrated that the local population of millisecond pulsars 
revealed by all-sky surveys at $\sim 0.4$~GHz
should be largely isotropic. This work, and early discoveries of two
recycled radio pulsars at high Galactic latitudes (Wolszczsan 1990)
inspired a number of 400~MHz pulsar surveys during the 1990s which led
to a sample of about 30 objects by the end of the decade. During that
time, studies of the scale height, velocity distribution and
luminosity function were performed (Lorimer 1995; Cordes \& Chernoff
1997; Lyne et al.~1998) and it was found that the local (within a few kpc)
millisecond pulsar population potentially observable was comparable in
size to the equivalent population of normal pulsars. One conclusion
from these studies is that the populations of millisecond and normal
pulsars are consistent with a single velocity distribution applied to
all neutron stars at birth (Tauris \& Bailes 1996).

\section{A new analysis of the millisecond pulsar population}

We are now in an era where significant further understanding
of the millisecond pulsar population should be possible in the
coming years. As a starting point, I present here a snapshot
analysis of the sample of millisecond pulsars detectable by
the Parkes multibeam surveys prior to the current high time-resolution
universe surveys (see Keith's contribution in this proceedings,
and Keith et al.~2010). The total number of millisecond
pulsars from these surveys now numbers 58. This number turns out
to have more-or-less asymptoted, but may still increase further
thanks to a number of new discoveries\footnote{Data analysis by
the Einstein@Home team has so far discovered 23 sources including
one highly dispersed millisecond pulsar, while Mickaliger et al.~(2012)
have recently announced the discovery of five further millisecond pulsars.}
by reanalyses of the Parkes multibeam survey of the Galactic plane.

Using the snapshot approach, I have
developed a model (hereafter referred to as model A)
which has the following parameters:
(i) a log-normal luminosity function with an identical mean
and standard deviation (i.e.~--1.1 and 0.9)
to that found by Faucher-Gigu\`ere \& Kaspi
(2006). This function was found to
be consistent with recycled pulsars in globular clusters recently
by Bagchi et al.~(2012); (ii) 
a manually-tweaked period distribution with a peak at
3~ms; (iii)
an exponential scale height with a mean of 500~pc; (iv)
a Gaussian radial distribution with a standard deviation
of 7.5~kpc. The period 
distribution was arrived at by initially choosing periods
from a distribution which is uniform in $\log P$ between 1 and 30~ms.
I adjusted the relative weighting of the bins to arrive at a distribution
which most closely matches the observed sample. The $z$ 
distribution was motivated by my earlier results (Lorimer 1995)
based on the low-frequency surveys.

\begin{figure}[hbt]
\begin{center}
 \includegraphics[width=\textwidth]{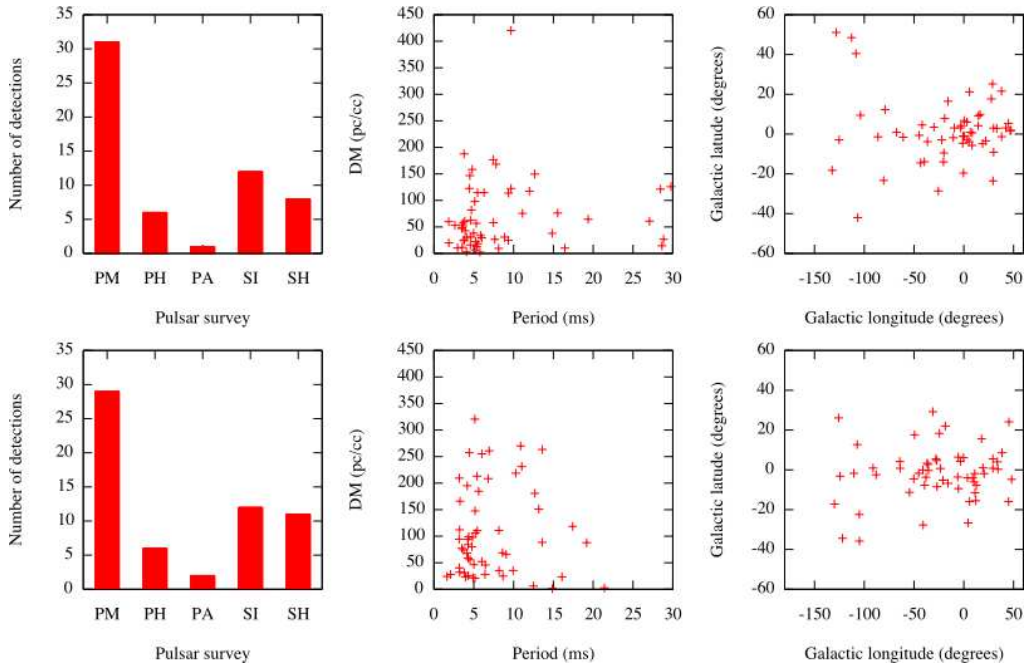} 
 \caption{The sample of millisecond pulsars detected in the five major
surveys (top panels) confronted with the equivalent distributions from
model A shown in the lower panels (see text).}
   \label{modelA}
\end{center}
\vspace{3mm}
\end{figure}

As can be seen in Fig.~1, this model provides a reasonable match
to the observed sample. I define reasonable in this context in terms
of a comparison of the observed and predicted survey yields and also
by looking at the observed distributions of spin period, $P$, dispersion
measure, DM, Galactic longitude, $l$, and Galactic latitude, $b$. In
Table 1, along with the total number of potentially 
observable\footnote{These are the number of pulsars in the model Galaxy
whose beams intersect our line of sight --- i.e.~uncorrected for beaming
effects.} I tabulate two figures of merit for this model: ${\cal Q_{\rm KS}}$
and $\chi^2_{\rm nobs}$. The former is the base 10 logarithm of the
product of the four individual Kolmogorov-Smirnoff (KS) tests between
$P$, DM, $l$ and $b$. The latter is the reduced $\chi^2$
computed from the observed and predicted numbers of pulsars.

\begin{table}
\begin{tabular}{rrrrr}
Model & Modification compared to model A & ${\cal Q_{\rm KS}}$ & $\chi^2_{\rm nobs}$ & $N_{\rm Galaxy}$ \\
\hline
A     & None              & --2.3             &  0.7       & 31,700 \\
B     & Mean of log-normal reduced to --1.2   & --2.1             &  1.3       & 36,000 \\
C     & Uniform surface density over the disk      &--7.1             &  2.8       &  2,670 \\
D     & Cordes \& Chernoff (1998) spin period distribution &--12.8            &  1.1       & 43,700 \\
E     & 100 pc vertical ($z$) scale height&--4.9             &  8.1       & 21,000 \\
F     & 1 kpc vertical ($z$) scale height &--3.1             &  1.3       & 40,800 \\
G     & 6.5 kpc radial ($R$) scale length&--2.6             &  0.5       & 30,800 \\
H     & 8.5 kpc radial ($R$) scale length&--3.4             &  0.9       & 31,800 \\
I     & Gaussian spin period distribution &--6.0             &  0.4       & 28,700 \\
\end{tabular}
\caption{Summary of the simulation results obtained from
our snapshot modeling of the millisecond pulsar population. 
From left to 
right we list the model, base-10 logarithm of the combined
Kolmogorov Smirnoff probability (${\cal Q_{\rm KS}}$), the reduced
chi-squared value ($\chi^2_{\rm nobs}$) and the number of
potentially observable millisecond pulsars in the Galaxy
($N_{\rm Galaxy}$).}
\end{table}

Also listed in the Table are the equivalent numbers for a number of
other models B--I. The approach I took here was to investigate the impact
of each assumption made in model A by changing it, but keeping all other
assumptions constant. In model B, to show the impact on the choice
of luminosity function on the results, I made only a small change to
mean of the log-normal function, i.e.~--1.1 to --1.2. The figures of
merit are comparable to model A, but the number of pulsars increases by
about 14\%. For model B, to show that the sample does require a radial
dependence on number density, I generated pulsars assuming constant
number density on the plane throughout the model galaxy. Here, both
figures of merit are substantially poorer due to the gain in number of
detections at higher Galactic longitudes. The number of pulsars required
in the model is substantially reduced, since it is now much easier to
detect the population at larger $R$. Models G and H, where the scale
length is varied by $\pm 1$~kpc from the value in model A show that
our ability to constrain the scale length is currently not very good.
In model D, I adopt the Cordes \&
Chernoff (1997) spin period distribution. Although the relative survey 
yields are satisfactory, this distribution predicts a substantial fraction
of millisecond pulsars with $P<2$~ms which is much higher than that observed.
A similar result is found using a Gaussian period distribution (model I) with
a mean of zero and a standard deviation of 10~ms\footnote{Only the positive
periods from this distribution are used for this model!}.
Reducing the scale height of the population in model E to 100~pc substantially
worsens the agreement with the observed data, while increasing the scale
height to 1~kpc (model F) has less of an effect. 

\section{Suggestions for further work}

The analysis presented here will be described further in a forthcoming
Parkes multibeam survey paper. It represents a first step towards a more
detailed understanding of the millisecond pulsar population in the
Galaxy. Further work is encouraged to account for the following
subtleties not included here. Of particular interest are studies of the
motion of millisecond pulsars in the $P-\dot{P}$ diagram and the 
relationship to the low-mass X-ray binary population. The work of Kiziltan
\& Thorsett (2010) and Tauris et al.~(2012) 
relates directly to the first issue. Further
work in this area, along the lines of the population syntheses carried
out by Story et al.~(2007) seem to be the next logical step. Significant
progress is now being made in modeling the binary evolutionary steps and
predicting distributions for orbital parameters for the binary population
(see, for example, Belszcynski et al.~2008). Combining all these
elements into an all-encompassing synthesis of the millisecond pulsar
population which accounts (as far as possible) for the observational
selection effects is now a major goal of future studies. On the road
to such a lofty goal, past experience with the normal pulsar population
(see, for example, Lorimer 2009) suggests that it will be extremely
profitable to break up the steps into a number of smaller problems.
One such example is the radio-selected sample of millisecond pulsars
revealed by {\it Fermi} (Ray et al.~2012). A careful study of the
selection effects impacting this sample should now be undertaken in
order to fully understand the impact of these discoveries on our
knowledge of the millisecond pulsar population.

\begin{discussion}

\discuss{Keane}{Pulsar surveys are not exhaustively searched. Is there
any means to account for this in your modeling?
}

\discuss{Lorimer}{There is a ``fractional completeness'' parameter
in {\scshape psrpop} that can be tweaked.}

\discuss{Ransom}{We heard earlier in the week for millisecond pulsars
that there might be some interesting spectral dependencies. How easy would
it be to add in the larger 350~MHz surveys into this modeling?}

\discuss{Lorimer}{Not too difficult. We just need the information describing
the sky coverage of these surveys. What you then need to do is to add
the pulsar spectra to your models, but that will hopefully teach you something
about that. [Note added in write-up: see the contribution by Youling You et al.
in these proceedings]}

\discuss{Heras}{For the 20\% of millisecond 
pulsars which are isolated, are there any
differences between this population and those that are members of binary
systems?}

\discuss{Lorimer}{As far as I am aware (and I looked at this last a few
years ago), there are no significant differences
in the population of isolated or binary millisecond pulsars in terms of
$P$, $L$, spatial distribution etc. What I think is interesting is whether
the binary population syntheses can match that 20\% isolated millisecond
fraction that we currently observe. That remains to be seen.}

\end{discussion}

\end{document}